\documentstyle[12pt,aaspp4]{article}

\lefthead{Secker}
\righthead{A Radial Color Gradient}

\begin{document}

\title{A Constraint On The Formation Of Dwarf Elliptical Galaxies In The Dense
Coma Cluster Core}

\author{Jeff Secker\altaffilmark{1}}

\affil{Washington State University, Program in Astronomy, Pullman, WA 99164-3113\\  secker@delta.math.wsu.edu}

\altaffiltext{1}{Based on observations taken at KPNO, which is operated 
by AURA, Inc.\ under contract to the National Science Foundation.} 

\begin{abstract}
Deep CCD photometry and positions for a new sample of $\simeq 250$
dwarf elliptical (dE) galaxies in the Coma cluster core are analyzed.
A significant color gradient is detected in their projected radial
distribution, given by $\Delta (B$--$R)/\Delta (\log R_{\rm cc}) =
-0.08\pm0.02$ mag.  Calibrating these dE galaxies against Galactic
globular clusters yields a corresponding metallicity gradient which
goes as $Z \propto R_{\rm cc}^{-0.29}$.  Simulations reveal that this
gradient in the projected radial distribution corresponds to a true
(three-dimensional) radial color gradient which goes as $\Delta
(B$--$R)/\Delta (\log r) \simeq -0.20$ mag, or $Z \propto r^{-0.69}$,
over the same radial range.  If this radial gradient is a primordial
one, it is consistent with a model in which the intracluster gas
exerted a significant confinement pressure, impeding the outflow of
supernovae-driven metal-rich gas from the young dE galaxies.
\end{abstract}


\section{INTRODUCTION} 

Early-type dwarf-elliptical (dE) galaxies are a class of diffuse, low
luminosity, low surface brightness galaxies with luminosities in the
range $-18 \lesssim M_B \lesssim -8$ mag, and colors typically near
$(B$--$R) \simeq 1.4$ mag.  They are characterized by smooth surface
brightness profiles, and the non-nucleated dEs are well described by a
single exponential or a modified-exponential profile (e.g., Ichikawa,
Wakamatsu \& Okamura 1986; Ferguson \& Binggeli 1994; Cellone, Forte
\& Geisler 1994).  Color gradients are observed in some dE galaxies,
yet there does not appear to be any preference for red or blue color
gradients (Caldwell \& Bothun 1987; Vader et al. 1988; Cellone et al.
1994; Durrell 1996).  Dwarf elliptical galaxies appear to
be dominated by extensive dark-matter halos, with a mass-to-light
ratio that varies as $M/L \propto L^{-0.4}$ (Ferguson \& Binggeli
1994), consistent with the models of Dekel and Silk (1986).  With a
typical mass on the order of $10^9 M_{\odot}$, dE galaxies may be the
evolved state of primeval protogalactic fragments, the first
self-gravitating systems formed in a CDM cosmology (e.g., Searle \&
Zinn 1978; Larson 1985).  While the formation of dE galaxies is not
completely understood, the review by Ferguson \& Binggeli (1994)
provides an excellent overview of this field.

In a recent simulation, Moore et al. (1996) illustrated that a large
fraction of a cluster's dE galaxy population can arise as remnants of
late-type galaxies, which are harrassed as they infall to the
cluster.  In these simulations, the primary infalling galaxy is
tidally disrupted and stripped by the mean cluster field and by
encounters with individual cluster galaxies. The result is one primary
dE galaxy (corresponding to the remnant core of the late-type galaxy),
with the possibility that additional bound stellar systems (presumably
dwarf galaxies) can form in the tidally-stripped galaxy debris.  In
comparison, traditional formation scenarios relevant to dE galaxies in
a cluster environment assume that they are primeval, and attribute
their observed characteristics to an interdependence between {\em
external} and {\em internal} factors.  For a dense environment such as
the Coma cluster, the dominant external factors would include: a
confinement pressure exerted by the intracluster medium, ram-pressure
stripping of galactic gas due to motions through the intergalactic
medium, and tidal destruction of low-mass dE galaxies in the
dense cluster core.  The internal factors relevant to dE galaxy
formation include extensive dark-matter halos, and an expulsion of
residual gas via supernovae-driven winds (e.g. Wyse \& Silk
1985; Dekel \& Silk 1986; Silk, Wyse \& Shields 1987; Babul \& Rees
1992; Marlowe et al. 1995; Babul \& Ferguson 1996).  The combined
result is expected to be a single major star formation epoch, a
stellar population similar in age and integrated light to globular
clusters, with a galaxy mass and metallicity dependent upon location
with respect to the cluster center.

The Coma cluster (e.g., Thompson \& Gregory 1993; Colless \& Dunn
1996) is the nearest of the very rich Abell clusters ($cz \simeq 7200$
km/sec, Abell richness class 2), and the core of this dense
environment provides an excellent location in which to study the
properties of dE galaxies.  An analysis of the photometric colors of
dE galaxies can provide valuable insights into the properties of
individual galaxies, and can also reveal statistical correlations for
populations of galaxies (Caldwell \& Bothun 1987; Evans, Davies \&
Phillipps, 1990; Garilli et al. 1992; Cellone et al.  1994).
In this Letter, the distribution of $(B$--$R)$ color for individual dE
galaxies is analyzed as a function of their clustercentric radius, and
a significant radial gradient in this color is detected.  This new
finding is then discussed in the context of dE galaxy formation and
evolution in a gas-rich, ultra-dense environment such as the Coma
cluster.

\section{ANALYSIS}

The data sample used here is a subset of the full data sample
presented in Secker (1995; 1996) and Secker \& Harris (1996=SH96).
Briefly, a $700$ arcmin$^2$ region of the Coma cluster core was
mosaic-imaged in two colors ($R$, $B$) with the KPNO 4m, in 1.1--1.3
arcsec seeing.  The three cluster fields considered here together span
a range in clustercentric radius of $1.33 \le R_{\rm cc} \le 23.33$
arcmin, with 1 arcmin $\simeq 20.9 h^{-1}$ kpc.  $R$-band magnitudes
were derived with Kron's $2r_1$ aperture (e.g., Bershady et al. 1994),
and corrected to true total magnitudes with simulations (Secker 1996;
1996).  Constant aperture magnitudes ($r_{\rm ap} \simeq 0.56h^{-1}$
kpc) were used to calculate accurate ($B$--$R$) colors, prior to which
the seeing was matched between the corresponding $R$ and $B$ images.
Our cluster membership is based upon the object's $(B$--$R)$ color:
the dE galaxy population is clearly evident on a $R$, $(B$--$R)$
color-magnitude diagram as a narrow sequence not present on the
control field (SH96).  As illustrated by Biviano et al. (1995) and
SH96, this technique enormously reduces contamination due to
background galaxies.  In this Letter, I restrict the analysis to
objects in the dE color range $1.05 \le (B$--$R) \le 1.6$ mag, while
in magnitude, the sample is restricted to the range $15.5 \le R \le
20.0$ mag.  (A value of $H_0 = 75$ gives a distance modulus of 34.9,
for which $R = 15.5$ mag corresponds to $M_R = -19.4$ mag, an upper
limit for dE galaxies.)  Above $R = 19.5$ mag, dE galaxies are clearly
resolved as nonstellar, and all stellar objects have been excluded.
In this color and magnitude range there are 340 objects on the program
field (dominated by cluster dE galaxies), compared to 87 objects
(scaled to the same area) on the control field.  At $R=20.0$ mag, the
sample is $\simeq 90$ percent complete in magnitude, and $\simeq 100$
percent complete in color, with a typical color uncertainty of
$\pm0.03$ mag.

In Figure \ref{fig1}, the ($B$--$R$) color for each of the 340
program-field objects is plotted versus the logarithmic projected
clustercentric radius, adopting NGC 4874 as the cluster center.
Superimposed upon the scatterplot of Figure \ref{fig1} are larger
solid circles representing the trimmed mean $(B$--$R)$ color in two
arcmin radial annuli, plotted at the geometric mean radius of the bin.
The uncertainty estimates on these points correspond to the standard
error of the mean.  Initially evident in Figure \ref{fig1} is the
trend for decreasing mean color with increasing radius, though
obscured to some degree by the scatter in the binned values.

\placefigure{fig1}

The solid line in Figure \ref{fig1} is the result of a weighted
least-squares fit to the mean ($B$--$R$) colors.  The slope and
intercept for this regression line are given by $\Delta (B$--$R)
/\Delta (\log R_{\rm cc}) = -0.08 \pm 0.02$ and $(B$--$R) =
1.46\pm0.02$ mag at $R_{\rm cc}=1$ arcmin.  This is a radial color
gradient, significant at the four sigma level, which corresponds to a
decrease of $0.10$ mag in the mean color of dE galaxies over the
radial range $1.33 \le R_{\rm cc} \le 23.33$ arcmin.  {\em Thus, we
detect a significant gradient in the distribution of galaxy color
versus projected clustercentric radius, which works in the sense that
dE galaxies at smaller clustercentric radii are redder in the mean.}

The $(B$--$R)$ color index, like $(B$--$I)$ and the Washington $(C$--$T_1)$
color, is a sensitive and accurate estimator of the total heavy
element abundance (metallicity) for old stellar populations (Geisler
\& Forte 1990; Couture, Harris \& Allwright 1991; Secker et al. 1995).
Several of the Local Group dwarfs show evidence for more than one
episode of star formation (e.g., Caldwell \& Bothun 1987; Sarajedini
\& Layden 1995; Smecker-Hane, Stetson \& Hesser 1995), and this may be 
true for a fraction of the early-type dE galaxies located in the
environment of rich clusters. However, it is a fair assumption that
the majority of these cluster dE galaxies are composed of a metal-poor
stellar population, similar in age to globular clusters (Ferguson \&
Binggeli 1994).  While both age and metallicity affect absorption
features in the spectra (and therefore the integrated color) for
systems of old stars, the effect of metal abundance dominates over age
effects in the observed $(B$--$R)$ color (Worthey 1994). As well,
changes in the integrated color induced by starbursts are short lived
(about $10^9$ years), after which the integrated colors return to
normal (Charlot \& Silk 1994).  {Thus in this sample of Coma cluster
dE galaxies, I assume that the redder dE galaxies are more metal rich
than the bluer dE galaxies}.

In order to quantify this metallicity scale, I adopt a
color-metallicity relationship derived from metallicities and
integrated $(B$--$R)_0$ colors for 82 Galactic globular clusters (from
Harris 1996), given by [Fe/H]$=(3.44\pm0.09)(B$--$R)_0-(5.35\pm0.10)$.
This color-metallicity relationship is calibrated over the range $0.85
\lesssim (B$--$R)_0 \lesssim 1.45$ mag, which I linearly extrapolate
to $(B$--$R)$ = 1.6 mag.  In this analysis, the observed $(B$--$R)$
colors are used, uncorrected for a foreground reddening of $E_{(B-R)}
\simeq 0.02$ mag.  Combined uncertainties in the photometric
calibration, reddening and the metallicity calibration give rise to
typical external uncertainties for [Fe/H] on the order of 0.2 dex.

The right-side axis of Figure \ref{fig1} labels this metallicity
scale, valid for the 340 program-field objects and the mean values in
the radial annuli.  In terms of metallicity, the slope and intercept
of the regression line are given by $\Delta$[Fe/H]$/ \Delta (\log
R_{\rm cc}) = -0.29\pm 0.07$ and [Fe/H] = $-0.32\pm0.02$ dex at $R=1$
arcmin.  This slope indicates a significant radial metallicity
gradient, which goes as $Z \propto R^{-0.29}$. {\em Thus, for the
population of bright dE galaxies in the Coma cluster core, dE galaxies
closer to the cluster center are more metal rich in the mean.}
Considering the extreme nature of the environment in which these dE
galaxies formed and have evolved, a metallicity near [Fe/H]$\simeq
-0.32$ dex for the most luminous dEs is not inconceivable.

\section{DISCUSSION}

\subsection{A Primordial Origin For The Radial Color Gradient}

Is this detected radial metallicity gradient a manifestation of
primordial differences in the dE galaxies (a result of their location
within the cluster), or is this gradient due to an effect of different
origin?  The primary factor to be considered is the presence
of intracluster dust.  In 1993, Ferguson examined the relation
between the Mg$_2$ index and $(B$--$V)$ colors for elliptical
galaxies:  he determined that there is a measurable quantity of dust
in the Coma intracluster medium, and estimated $E_{(B-V)}
\lesssim 0.05$ mag, such that $E_{(B-R)} \lesssim 0.08$ mag.
Via simulations, he determined that a King model with a core radius of
0.5 Mpc was plausible for the dust distribution.  Assuming $E_{(B-R)}
\simeq 0.08$ mag at the cluster center, one would estimate a total
$\Delta E_{(B-R)} \lesssim 0.04$ mag out to the core radius.  Our data
spans nearly the same radial range, and from the analysis of the
projected radial color distribution we estimate $\Delta (B$--$R) =
0.10$ mag.  Thus, while the intracluster dust may contribute
significantly to our radial color gradient, it cannot be the sole
cause of the observed color gradient.

As detected by SH96, there exists a strong color-luminosity
correlation in this same data set, which works in the sense that the
more luminous dE galaxies are redder (in the mean); the detected color
gradient therefore also corresponds to a luminosity gradient.  Thus,
we may question whether any mass-dependent dynamical relaxation of the
dE galaxy orbits could have occurred.  For dE galaxies in a rich
cluster environment, both two-body relaxation and dynamical friction
occur over a timescale much greater than the Hubble time (Bahcall
1977; Merritt 1985).  West \& Richstone (1988) proposed that in a
cluster containing dark matter, violent relaxation occurring during
cluster collapse can result in spatial segregation between the luminous
galaxies and the dark matter.  However, their simulations show no
clear evidence for segregation between galaxies of the smallest and
intermediate masses. Thus the color distribution which we observe must
be primordial in nature.

\subsection{Modeling the Three-Dimensional Metallicity Gradient}

Figure 1 shows dE galaxy color plotted against {\em projected
clustercentric radius}.  Assuming that there is a color/metallicity
gradient in three dimensions for a centrally-concentrated galaxy
cluster, what is the effect of projection to two dimensions?
Specifically, projection will introduce significant scatter about the
mean color, with this scatter increasing towards the cluster center.
Since the mean value of the projected distribution involves a
summation along the line of sight, this increased scatter skews the
mean projected color in a biased manner, which results in a measured
color gradient which is shallower that the true gradient.

It is useful to explore simulated galaxy clusters for which the galaxy
color varies with the true (i.e., three-dimensional) clustercentric
radius $r$, and by doing so determine what true color distribution
gives rise to the observed radial color gradient in the projected
distribution.  The simulations performed here are based upon
spherically-symmetric King model galaxy clusters, with the true
spatial distribution of cluster galaxies over a radial range of $0 \le
r \le 3$ Mpc described by $\rho(r) = \rho_0 (1+r^2/R_c^2)^{-3/2}$,
with $R_c = 18$ arcmin.  The colors assigned to cluster members were
based solely upon their true radius $r$, as described below.  To this
was added a population of uniformly-distributed (control-field)
objects, whose colors were selected randomly and uniformly throughout
the interval considered here.

This simulated galaxy cluster and background population were then
projected to two dimensions (Figure 2), such that within the radial
range being considered, there were 255 cluster galaxies (small solid
circles) and 72 control-field objects (crosses).  Then, as was done
for Figure 1, the distribution of mean object color as a function of
projected clustercentric radii were analyzed for the sample of 327
simulated objects.  For the model illustrated in Figure 2, the color
distribution is given by $(B$--$R) = 1.64 - 0.20 \log(r) -
N(0,0.05)$.  The last term represents the model's scatter in color at
any $r$; it is normally distributed with a standard deviation of 0.05
mag.  To obtain the closest match to the observed distribution,
the true color distribution was varied between simulations; the
slope and intercept of the least-squares regression line in Figure 2
are equal to the values obtained for the dE galaxy sample of Section 2
(Figure 1).  While this color distribution function is a reasonable
one, it is not necessarily unique, and a different set of input
parameters may well yield a similar result for the projected radial
gradient.

While the distribution of the simulated objects in Figure 2 is not an
exact match to the observed objects in Figure 1, the simulated cluster
set is sufficiently similar to reveal the following about the nature
of the dE galaxy population in the Coma cluster.  First, the slope of
the projected radial color gradient, $\Delta (B$--$R)/\Delta (\log
R_{\rm cc}) = -0.08\pm 0.02$, corresponds to a much steeper slope in
the true galaxy distribution, $\Delta (B$--$R)/\Delta (\log r) \simeq
-0.20$, or $Z \propto r^{-0.69}$.  Second, the total color
change of $\Delta (B$--$R) = 0.10$ mag for the projected distribution
corresponds to a color change of $\Delta (B$--$R) \simeq 0.25$ mag,
or $\Delta$[Fe/H]$ \simeq 0.86$ dex, in the true distribution, over the
(same) observed radial range.

While our simulated cluster reproduces the main features of the
observed dE galaxy color-projected radius distribution very well,
there are noticeable differences between Figures 1 and 2.  Although
the dominate effect is due to projection, there are secondary effects
which will add further second-order scatter: deviations of the Coma
cluster from spherical symmetry, motions of individual galaxies from
their primeval location, variations in the level of preenrichment
between galaxy cluster subclumps, and variations in the density of the
intracluster medium (see the discussion below).  These secondary
effects were not included in our simulation.

\subsection{Interpretation}

Any successful model for the formation of dE galaxies in a cluster
environment, regardless of the mechanism or epoch, must naturally
account for the radial metallicity gradient described in the previous
section.  The ``late- accretion'' model of Silk, Wyse \& Shields
(1987=SWS87) postulates that the intracluster gas was enriched by the
metal-rich gas outflows from dEs at an early epoch (i.e., $10 \gtrsim
z \gtrsim 5$).  Then during the epoch of cluster formation, this
diffuse gas is compressed and cools, subsequently infalling and
accreting onto the halos of the most massive dwarf galaxies; vigorous
starbursts can then follow.  SWS87 assume that the star formation
within a dwarf galaxy is governed by the gas accretion rate, which is
regulated by the proximity of the dwarf galaxy to a more massive
companion; i.e., the tidal field of a large central galaxy effectively
strips gas from the nearest dwarf galaxies, such that the star
formation rate decreases towards the cluster center.

The SWS87 model is applicable to dwarf galaxies in a cluster
environment, and it describes a transition from dE galaxies to dIr
galaxies during the present epoch of galaxy cluster formation.  They
also postulate that this accretion mechanism could supply the gas
necessary for star formation in the nucleus of the more massive dE
galaxies, thus explaining the presence of a nucleus in the brighter dE
galaxies.  If this is indeed the case, an observable metallicity
gradient within the dE,N galaxy population would ensue, with those
further from the cluster center being more metal rich in the mean.
The sample of 340 objects that are analyzed above includes all
detected dE galaxies in the absolute magnitude range $-19 \lesssim M_R
\lesssim -14.5$ mag. Within this magnitude range, and within our
sample, we expect the fraction N(dE,N)/N(dE) to vary from nearly 100
percent for the brightest down to 20 or 30 percent for the fainter
dwarfs (van den Bergh 1986); that is, the sample of bright dE galaxies
will be dominated by nucleated dE galaxies.  If these dE,N galaxies
result from this late accretion, SWS87 predict that a positive radial
metallicity gradient would be observed, which is not consistent with
the radial metallicity gradient we detect.

The model introduced by Babul \& Rees (1992=BR92) naturally explains
the observed negative metallicity gradient.  BR92 postulate that the
intracluster medium (ICM) played a dominant role in the evolution of
dwarf galaxies, as an external pressure which confined the
supernova-driven gas, preventing gas loss in a way that the dark halos
themselves could not.  BR92 postulate that if this ICM pressure dominated
over the counter-effect of ram-pressure stripping, it could halt the
outflow of metal-rich gas from the young dEs.  This metal-rich gas
would then cool and fall back onto the galaxy core; this gas would
then be available for subsequent star bursts.  They attribute the
paucity of field dE galaxies at the present time to the lack of
confinement pressure; the field dE galaxies more easily expelled their
gas, inhibiting further star formation, and have subsequently faded
away.  In fact, BR92 and Babul \& Ferguson (1996) have proposed
that the faint blue galaxies (e.g., Koo \& Kron 1992) are a field
population of star-bursting dE galaxies at redshifts $0.5 \lesssim z
\lesssim 1$, which will subsequently fade.

This ICM is an integral component of rich galaxy clusters like Coma,
and while it is diffuse (i.e., $n \simeq 10^{-2}$--$10^{-3}$ cm$^{-3}$ in the
core), the total mass of the ICM is comparable to the total mass in
luminous matter throughout the cluster.  With a temperature near $T
\simeq 10^7$--$10^8$ K (Fabian et al. 1994), the ICM pressure in the
cluster core is on the order of $(nT)_{\rm ICM} \gtrsim 10^4$
cm$^{-3}$ K, sufficient to confine the outflowing gas within the
dark-matter halos of the dE galaxies.  Thus the dense core of the
Coma cluster provides an excellent laboratory in which to study the
effects of variation in the density of the ICM on dE galaxy
metallicity.  Following the logic of BR92, and since the ICM density
peaks in the cluster core near NGC 4874 (White, Briel \& Henry 1993),
the dEs nearest to the cluster center should be on average both {\em
more massive} and {\em more metal rich}, than dEs farther from the
cluster core.  Thus the detection of a negative radial metallicity
gradient within the dE galaxy population of the Coma cluster core is
consistent with the model of BR92.

\acknowledgments

This research was supported by an NSERC operating grant to W.E. Harris,
by the Ontario Ministry of Colleges and Universities (a 1994/95
scholarship to the author), and by the Department of Physics and
Astronomy at McMaster University.  I would like to thank Arif Babul,
Pat Durrell, Doug Geisler, Bill Harris and Dean McLaughlin for their
helpful comments and discussions.

\clearpage

\figcaption[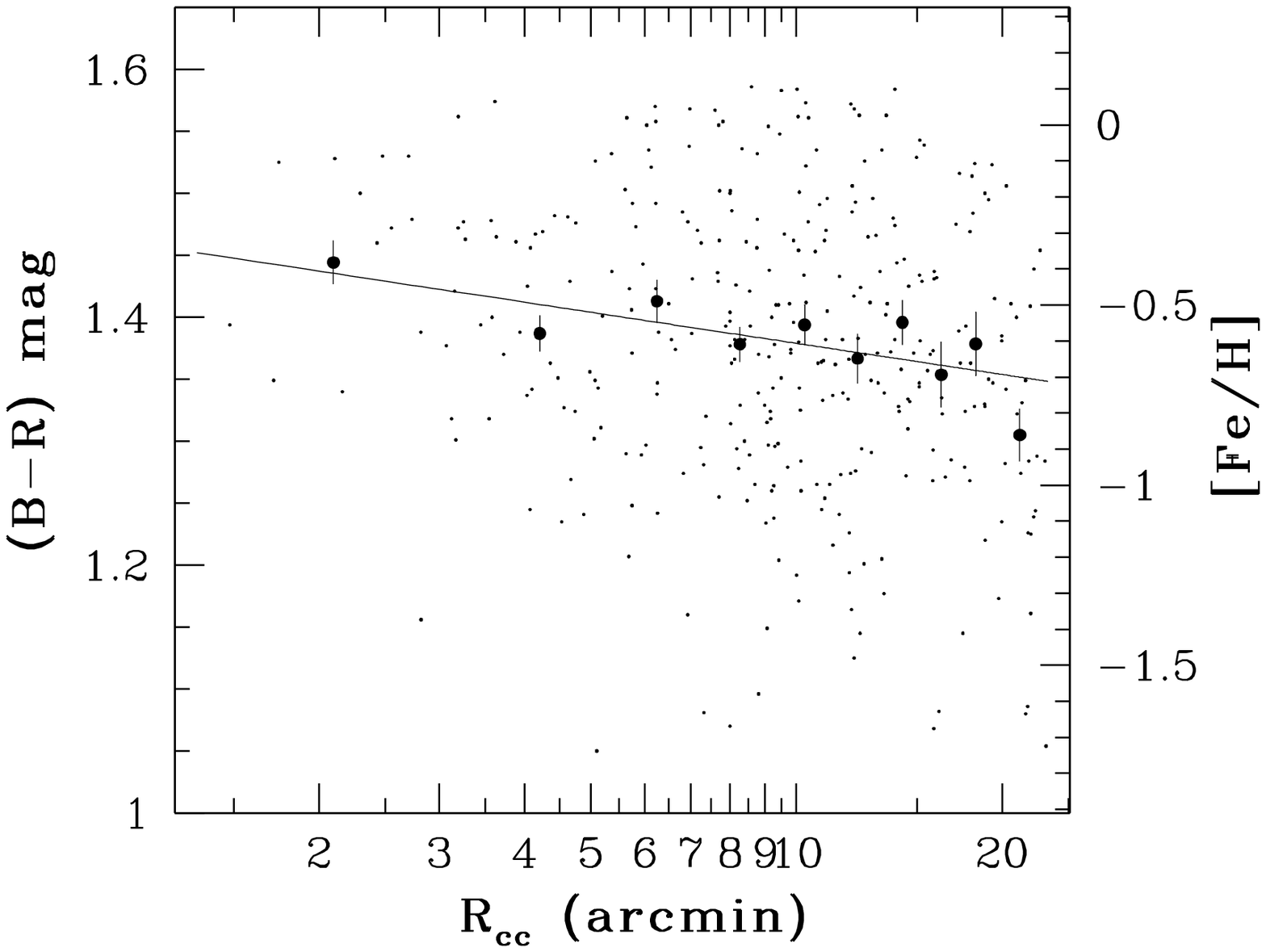]{Color/metallicity versus the logarithmic projected 
clustercentric radius for a sample of 340 dwarf elliptical (dE) galaxy
candidates with $15.5 \le R \le 20.0$ mag and $1.05 \le (B$--$R) \le
1.6$ mag.  The small points represent the individual dE galaxies,
while the larger solid circles represent the trimmed mean $(B$--$R)$
color (and standard error of the mean uncertainties) within
radial annuli.  The weighted least-squares regression line is shown
through these points; its slope yields a radial color/metallicity
gradient, significant at the four sigma level. \label{fig1}}

\figcaption[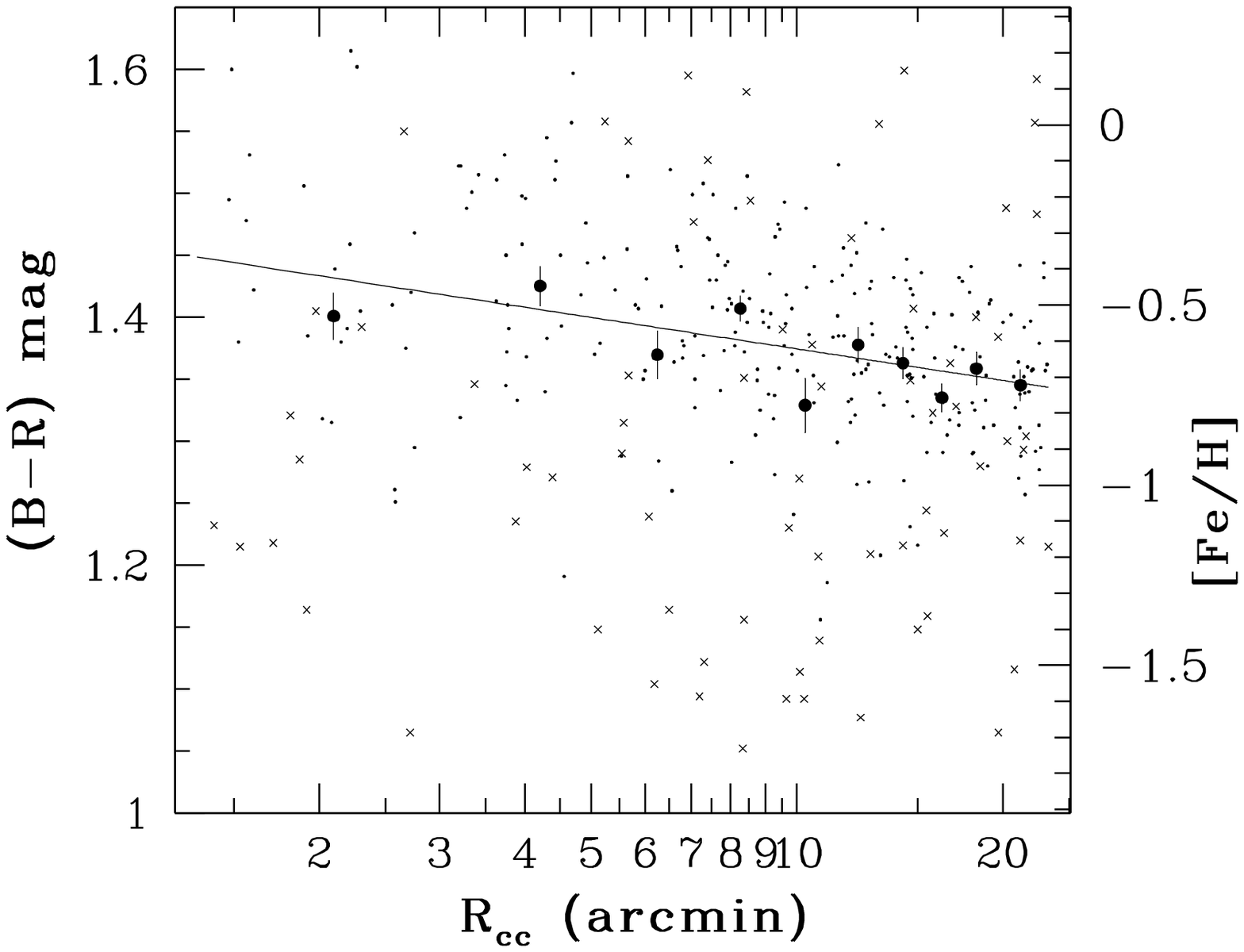]{A model for the true spatial and color distribution 
which is consistent with the projected radial color distribution of
Figure 1.  Plotted here is a sample of objects drawn from a simulated
galaxy cluster plus a uniform background population of galaxies.  The
small points represent the individual cluster galaxies, the crosses
represent background galaxies, and the larger solid circles represent
the trimmed mean $(B$--$R)$ color of all objects within radial annuli.
The least-squares regression line for this projected distribution is
equal to that obtained for Figure 1. \label{fig2}}

\clearpage 

\plotone{RCDfig1.eps} 

\clearpage 

\plotone{RCDfig2.eps} 

\end{document}